# Emergence of online communities: Empirical evidence and theory

**Authors:** Yaniv Dover[1*] and Guy Kelman[2]

**Affiliations:**

[1] Jerusalem School of Business Administration, Hebrew University, Jerusalem, Israel 91905.

[2] School of Computer Science, Hebrew University, Jerusalem, Israel 91905.

*Correspondence to: yaniv.dover@mail.huji.ac.il.

Online communities, which have become an integral part of the day-to-day life of people and organizations, exhibit much diversity in both size and activity level; some communities grow to a massive scale and thrive, whereas others remain small, and even wither. In spite of the important role of these proliferating communities, there is limited empirical evidence that identifies the dominant factors underlying their dynamics. Using data collected from seven large online platforms, we observe a universal relationship between online community size and its activity: First, three distinct activity regimes exist, one of low-activity and two of high-activity. Further, we find a sharp activity phase transition at a critical community size that marks the shift between the first and the second regime. Essentially, it is around this critical size that sustainable interactive communities emerge. Finally, above a higher characteristic size, community activity reaches and remains at a constant and high level to form the third regime. We propose that the sharp activity phase transition and the regime structure stem from the branching property of online interactions. Branching results in the emergence of multiplicative growth of the interactions above certain community sizes.

**Introduction**

Peer-to-peer group interactions are prevalent in online platforms. People regularly participate in online groups and communities, interact with other members, and are affected by their peers[1,2]. Still, there is little empirical evidence that pins down the factors that determine whether a community will keep thriving with activity or fail to attract or retain active members. The extant literature discusses several factors that are important in maintaining meaningful social group action[3-7]. First and foremost is the number of committed group members at a given time. This is a prominent indicator of an active community, even if the commitment levels are heterogeneous[3]. The second factor is the minimal level of interdependence required between group members to induce any interaction within the community. Third, the marginal returns on contribution should be non-decreasing[3]. Other factors such as group context and social network structure have also been surveyed in the theoretical literature, but some studies[6] suggest that these effects are "second order." Here, we wish to gain empirical insight into online communities' stability by investigating the relationship between activity and size. It is not immediately clear from the literature what the expected exact activity–size relationship should be although some studies suggest that this relationship should strongly depend on the underlying production function, context, competition[8-9] and heterogeneity[3]. To investigate this, we collected and analyzed several rich datasets that contain hundreds of thousands of online communities, spanning a time frame of more than a decade.

**Results**

Figure 1 shows the median per-capita activity of online communities as a function of their size for a large website that hosts thematic discussion-based online communities (TAP dataset, see Methods for details). On this platform, users can choose to either initiate or terminate their own communities, which allows us to observe their "organic" life cycle across time. We use per-capita activity, i.e., mean user activity, to control for linear size effects. Figure 1 shows three distinct size regimes of activity which can be summarized as follows:

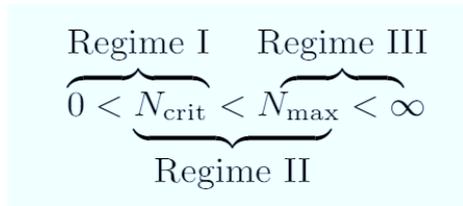

*Regime I* spans small communities of up to about 20 members. In this regime, activity is sporadic, and, across the lifetime of a community, the mean number of posts per user is low, at around three posts per user. The slope of the dependence of mean activity on community size is 0.086 (with a std. error of 0.018). Analysis shows, in this regime, a community requires more than 12 new users just to "encourage" community members to increase their posting rate by one additional message. Thus, the effect of community size on participation is very small. A sharp transition to *Regime II* occurs in communities of about 20 members, and ranges up to group sizes of about 50. This type of sharp transition around a critical mass, which we denote here by $N_{crit}$, is theorized in the literature[10], but empirical evidence for actual relevant examples are scarce. Below, in the Model section, we explain how this sharp transition likely results from the branching property of discussion trees. The activity–size slope within Regime II is 0.91 (SE 0.085), meaning that it only takes one additional user to the community to be associated with an increase of one additional message to the posting rate of the typical user. This effect is an order of magnitude greater than in Regime I. In other words, community-wide effects on participation emerge in Regime II.

The transition into a third regime, *Regime III*, takes place in communities that roughly number 50 members or more. Here, like in Regime I, the slope is very small (0.022, SE 0.002). It seems, therefore, that there is a cap on community effects above a certain size, which we denote by $N_{max}$. In the Model section, we estimate $N_{max}$ and $N_{crit}$ among other parameters.

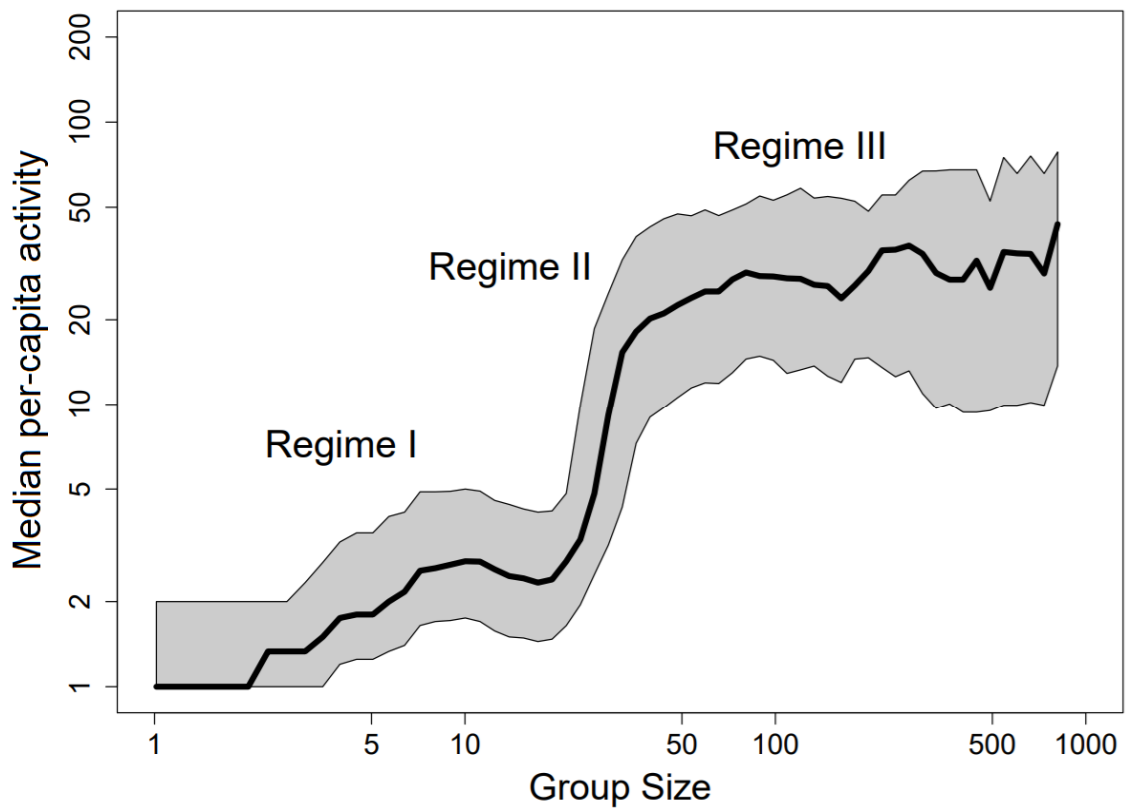

**Figure 1 | Mean activity across a range of community sizes in the TAP ("Tapuz Communes") data.** The thick solid line visualizes the median activity across community-size bins. The shaded areas mark the regions between the 25th to 75th percentiles.

Notably, Figure 1 summarizes the dynamics of communities over their entire lifetime. A lifetime in these data can span a few days or a few years, depending on activity levels and the time point at which activity ceases. In order to rule out a scenario where the three-regime structure is an artifact of some complex long-term dynamics, Figure 2 shows the same empirical dependency when we restrict the data to only the first year of the activity of each community since its inception. The pattern in Figure 2 is qualitatively identical to that in Figure 1.

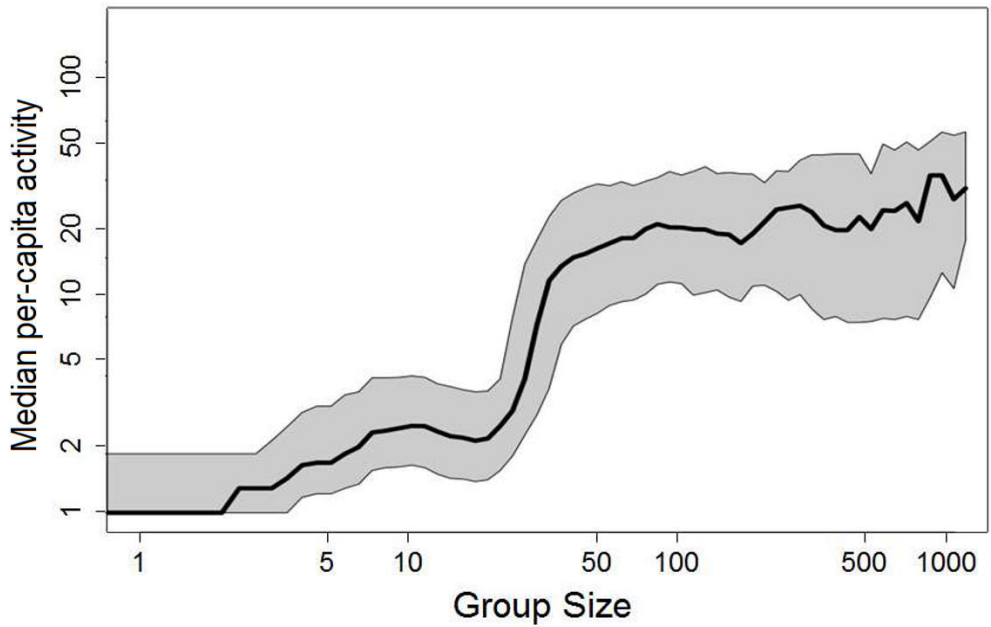

**Figure 2 | Mean activity across a range of community sizes in the TAP data, but only for the first year in the community's existence.** Curves and markings are similar to Figure 1.

Online peer-to-peer interactions can be thought of as trees of messages and replies. In these trees, messages are nodes and are connected by links that represent which message was addressed as a reply to which other previous message. This discussion tree begins with an initial seed message posted by a user. Other users can then post a reply to the seed message, i.e., link their messages to the initial one, creating a two-level tree. This tree can branch out further with replies to the replies at deeper levels, and so on. As we outline in the Model section, the rate of growth of a discussion tree depends solely on the distribution of the number of offspring, or replies to each message. If the mean number of replies per message is higher than one, the tree grows multiplicatively. If the mean number of replies is lower than one, the growth of the tree effectively decays geometrically. Therefore, we expect a sharp phase transition of discussion tree sizes, or number of messages, to occur as the mean response rate increases linearly within a community. We argue that the phase transition observed in Figure 1, across communities, stems from this branching property of peer-to-peer discussion trees. As a community grows in size, response rates also grow, and above the critical point of one reply per post, a sharp phase transition of multiplicative tree growth occurs. Evidence that this is indeed the actual scenario can be seen in Figure 3. This figure sketches typical discussion trees sampled from the data and arranged by community size. Each displayed tree is representative of the median depth at the given community size in the TAP dataset. The figure shows that the activity phase transition between Regimes I and II is strongly correlated with a sharp increase of tree depth. The offspring or response rate is also displayed in Figure 3 in the form of color-coding. Consistent with our theory (see Model section), the figure does show that the transition occurs around the offspring rate of unity.

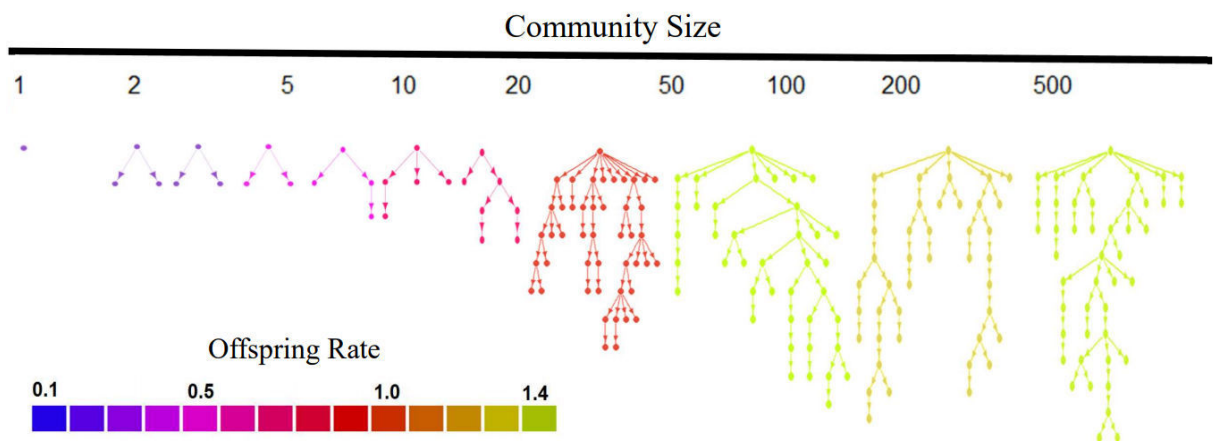

**Figure 3 | A schematic of selected discussion trees ordered by community size.** The nodes mark response posts and the root is the initial "seed" post. Links between nodes are the association of post and reply. The illustrated trees are color-coded according to the mean rate of replies to messages (mean offspring rate).

An outcome of the suggested theory is that the community's critical size, $N_{crit}$, strongly depends on community responsiveness. A crude but useful approximation for this relationship comes out under simplifying assumptions: $N_{crit} \cong \frac{1}{q}$ where $q$ is the community mean level of responsiveness (see the Model section). Our data are at the individual level, so it is possible to estimate the responsiveness $q$ for sets of communities and investigate the dependence of $N_{crit}$ on $q$. First, the analysis shows that the most common responsiveness in the TAP data is $q \cong 0.04$. This translates to an estimated $N_{crit} = 25$, which is roughly the actual transition point observed in Figure 1. Another test of the branching message tree theory is whether communities with higher responsiveness rates will correspond to lower critical community sizes. Unfortunately, an inherent property of these data is that the subsets of communities with homogeneous $q$s are small. This limits our ability to accurately validate the relationship $N_{crit} \cong \frac{1}{q}$, but it is still possible to test whether $N_{crit}$ decreases with increasing $q$, as we expect. To do this, we divide the data into four equal-count bins representing the quartiles of $q$ values, from low to high. Figure 4 shows the mean activity versus size curves for each of the four quartiles. The figure indicates that a sharp transition does exist for each subgroup and that the Regime I–Regime II transition shifts continuously leftward ($N_{crit}$ decreases) as the within-bin $q$ increases. For comparison, the expected $N_{crit}$ values that were calculated using the mean $q$ within each bin are: 1343.7, 37.7, 15.6, and 5.2, respectively for the first, second, third, and fourth $q$ bins. These estimations seem to be roughly consistent with the order of magnitude of the transition points in Figure 4. This is encouraging given the within-bin sample size limitation, heterogeneity within bins, and the crude approximation. Calculations at higher resolution, i.e., with 10 bins of $q$ (deciles) are shown in the inset of Figure 5 and support the expected reciprocal relationship between $N_{crit}$ and $q$.

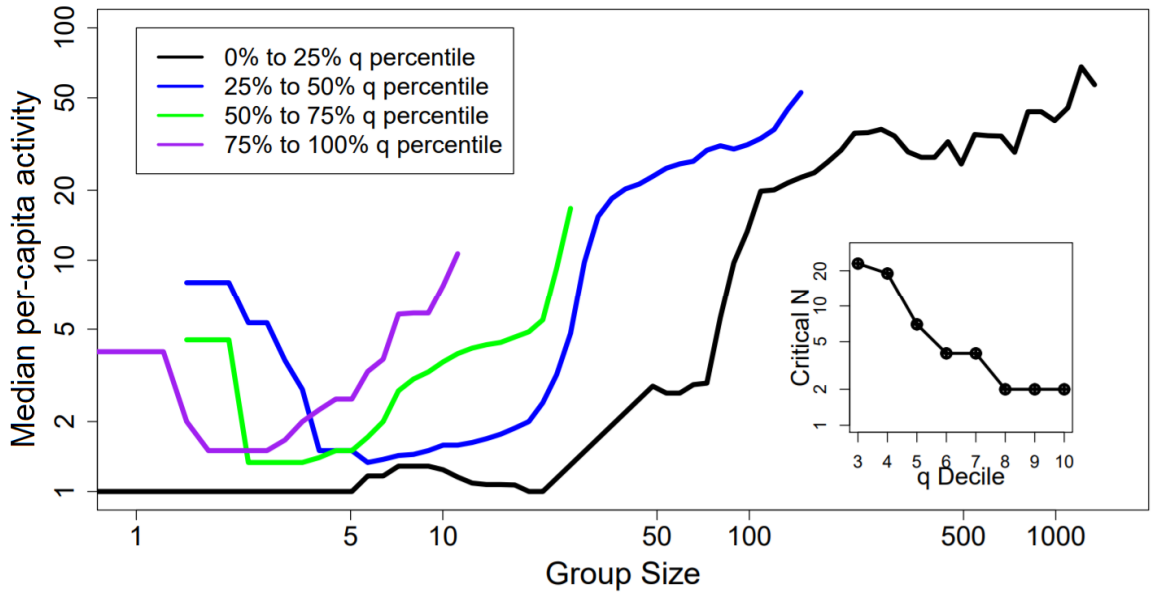

**Figure 4 | Activity median as a function of community size for communities grouped in four quartiles of response rate q (see legend for details).** The inset shows, for a tenfold partition of q values (deciles) the $N_{crit}$ as a function of within-decile $q$.

To test whether the three-regime pattern is unique to the TAP platform or is a more general phenomenon, we collected data from six additional online platforms. For consistency, we chose platforms that enable users to post messages and replies within distinct predefined communities (see the Methods section). The activity-size profiles for the additional platforms are laid out in Figure 5 and the supplement. Panel 5(a) illustrates the median activity as a function of community size for BRDS, the public dataset of Boards.ie. Much like the TAP set, BRDS has a distinct three-regime pattern. In this case, however, $N_{crit}$ is roughly 100, suggesting that the average community's ambient responsiveness is smaller on this platform, around $\cong \frac{1}{100}$. Panel 5(b) displays comment discussions from 8,446 random YouTube video pages. Note that the maximum point of the sharp transition for communities larger than $N_{max}$ is not as high as it is for TAP and BRDS. This is likely because YouTube is not specifically used for lengthy and continuous peer-to-peer discussions. The graph and patterns for HI5 and RED (Figures S1 and S2) are similar to the YOUT case, and are shown in the Supporting Information (SI). Next, panel 5(c) shows user discussions in a random collection of 21,000 Wikipedia Talk pages. The regime arrangement in WIKI is apparently missing Regime III, in contrast to TAP, BRDS, and YOUT. We believe that this may be a result of community size limitations, as the number of active concurrent authors per single title is normally small (median $N_{authors}$ < 10)[11], and substitution often supersedes growth. The transition between Regime I and Regime II occurs in communities larger than three authors, suggesting high responsiveness. Lastly, panel 5(d) shows the activity profile of 8,040 Goodreads discussion communities randomly collected and analyzed. Although Regimes II and III are distinguishable in this panel, Regime I is missing. We believe that this is the result of prevalent high levels of responsiveness in Goodreads; community members tend to be highly responsive to each other, even within small communities. It is important to note that our branching model, outlined in the next section, inherently accommodates the scenario in which Regime I or Regime III are not observed. For example, when $N_{crit}$ is very low, e.g., around a value of two, the transition to Regime II will occur in these very small communities and the first regime will not be discernible. Low values for critical size are expected when responsiveness is very high, i.e., within tightly-knit or highly interactive communities. Occurrence of Regime III depends on whether communities reach sizes of the order of $N_{max}$ or grow beyond them. If, however, we only observe communities smaller than $N_{max}$, Regime III will not exist, as we suspect occurs in the Wikipedia talk pages case.

In summary, the full three-regime structure is observed in five of the seven platforms: TAP, BRDS, YOUT, RED, and HI5. In the other two platforms, WIKI and GOODR, we only observe two regimes. We speculate that the differences stem from differing platform contexts and sampling constraints. However, the effect of platform context on the activity–size relationship is out of scope for this paper.

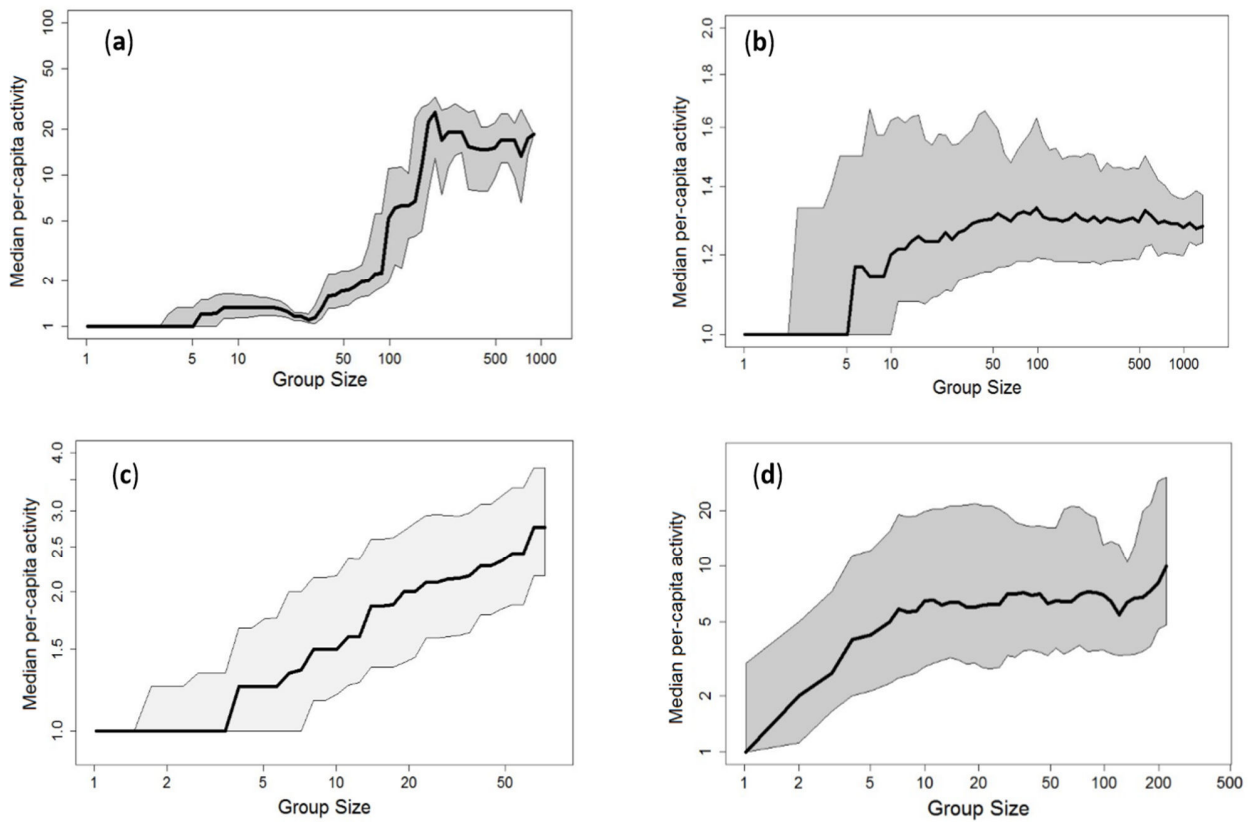

**Figure 5 | Activity median as a function of community size for other platforms.** (a) Boards.ie (BRDS), (b) YouTube (YOUT), (c) Wikipedia (WIKI), and (d) Goodreads

(GOODR). For visualizations of the two other platforms, see Figures S1 and S2 in the Supplementary Information.

**The Model**

We use a branching process model[12] to explain the observed activity–size patterns in which the three-regime structure is exhibited. In the model, a community consists of $N$ interacting members that generate trees of messages and their responses (e.g., see Fig. 1). We denote by $q_{i,j,k}$ the rate of $i$'s response to user $j$ for a given message $k$. In the basic scenario, both $q_{i,j,k}$ and $N$ are time-independent. We assume that across a given time period, there is a constant probability $p_i$ that user $i$ will post a seed message. A seed message is a tree made of one message that may later evolve, or branch, into a tree of more than one message. To further simplify, we assume that the propensity of user $i$ to respond to user $j$ does not depend on the posted message, i.e., $q_{i,j,k} = q_{i,j}$. The responsiveness in a community is then a matrix of response rates between all members, $Q = q_{i,j}$.

Evolution of message trees is modeled here as a Galton-Watson branching process. Let $Y_k$ be an i.i.d. random variable representing the number of replies that post $k$ receives. The offspring distribution $\phi(Y = \kappa)$ is the distribution of these replies, and we assume it to be homogeneous across users and time. In this scenario, $\phi(\kappa)$ depends only on $N$ and $Q$, i.e., $\phi(\kappa) = \phi(\kappa \mid N, Q)$. A useful quantity is the expected number of replies per post, $\langle \phi(\kappa) \rangle = \mu(N, Q)$. The general expression for the total number of messages $Z_{g,m}$ at tree depth $g$ in tree $m$ is given by the following iterative relation:

$$Z_{g,m} = \sum_{i=1}^{Z_{g-1,m}} Y_i \quad (1)$$

Where $Z_{0,m} = 1$ initially because discussion trees initiate with one message. Now, we denote by $\Gamma = \Gamma(s, g_{max} | N, Q, \phi(\kappa))$ the probability to observe a tree of size s and maximal depth gmax. Using (1), we write the expected number of replies per message $M_g$ at depth or generation g as the recursive relationship:

$$\langle Z_g \rangle \equiv M_g = \mu(N,Q) \cdot M_{g-1} \quad (2)$$

Since again, $M_0 = 1$, we arrive at the closed expression:

$$M_g = \mu(N,Q) \cdot g \quad (3)$$

Equation (3) demonstrates that the mean size of the tree is "geometrically sensitive" to the first moment of the distribution of replies, $\mu(N,Q)$. The critical point of growth occurs for $\mu(N,Q) = 1$. If $\mu(N,Q) > 1$, a super-critical branching process is in effect, and so tree posts will geometrically grow across generations. On the other hand, in the sub-critical case, $\mu(N,Q) < 1$, the expected number of replies shrinks geometrically. Finally, the mean community activity is the total activity in a community scaled to community size. In other words, it is the sum over the realizations of the tree sizes, i.e., the random variable $\gamma_s$ drawn from the distribution $\Gamma$ divided by community size:

$$A_{percap} = \frac{1}{N} \sum_{i=1}^{N \cdot p} \gamma_{s,t} \quad (4)$$

Here, we assumed a homogeneous probability p for all members. The size of trees in (4) is missing one realistic component, that the growth rate naturally decays with tree depth. Online discussions mature and saturate as the discussion tree grows. To account for this, we introduce a dependency of responsiveness on generation, $Q = Q(g)$ (see the SI for details). An analytic solution to (4) is intractable for our relevant case[13] for the same reasons that apply to equation (1). Therefore, in order to fit the data to our model and to see whether the model replicates the observed three-regime structure, we use numerical simulations and maximum likelihood estimations. From (3), the condition for a message tree's critical growth is:

$$\mu(N,Q) = 1 \quad (5)$$

$N_{crit}$ can be calculated using (5). A useful approximation that demonstrates the reciprocal relation between the critical size and community responsiveness arises if we assume $\phi(\kappa)$ to be binomial with a constant across-community responsiveness parameter $q$ (i.e., $Q_{i,j}(g) = q, \forall i,j$). Putting aside, for the moment, the dependence of responsiveness on generation, the expression for $N_{crit}$ is then:

$$N_{crit} = \frac{1}{q} \quad (6)$$

Essentially, (6) shows that the higher the inherent responsiveness of the community is, the lower the threshold.

We use a Maximum Likelihood Estimator (MLE) to test the agreement of the model with the data, and to find the best fit given these data. We use the TAP data, where we have the best user- and message-level resolution (see the Methods and SI sections for complete details). The MLE renders a statistical estimation of the offspring distribution parameters $\phi(\kappa | N, Q(g))$ – namely, the responsiveness matrix, $Q(g)$, $N_{max}$, and a third parameter $\lambda$, that represents the decline with tree depth of the probability that the tree will keep growing. For simplicity, we choose responsiveness for any $i,j$ to be: $Q_{i,j}(g) = q \cdot f(\lambda, g)$, where $q$ is constant, and the growth decay enters in the expression $f(\lambda, g)$. The functional form of $f()$ that best fits the data is $f_\lambda(g) = f(\lambda, g) = g^{-\lambda}$. The MLE results are given in Table 1, and the visualization of the fit is shown in Figure 6.

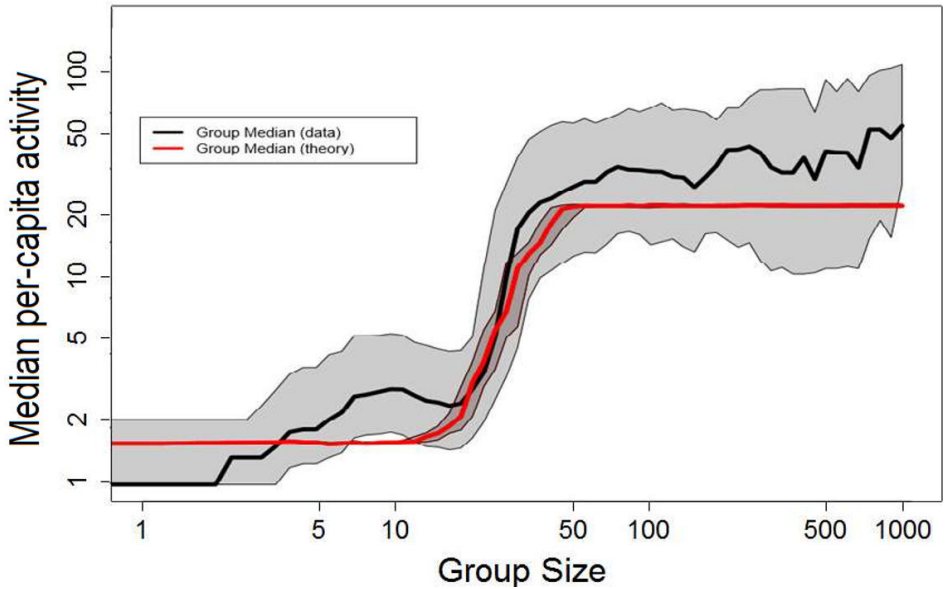

**Figure 6 | Fit of MLE estimations.** The model fit is shown in red and the respective percentile envelope overlay the data, as in Figure 1.

The red curve in Figure 6 is the result of the numerically simulated model of eq. (4) using the parameter values from Table 1. The red model curve reflects the structure of three regimes and the sharp Regime I–II transition. Note that the curve is within the 25%–75% percentile envelope of the observations, which is consistent with the encouraging fit measures (Table 1). Interestingly, the estimated responsiveness, $\hat{q} = 0.026$, is close to our initial, cruder estimations in the Introduction section and corresponds to $N_{crit} \approx 38$. Finally, another interesting observation is that, given the estimated value of $\lambda$, the functional form of the rate of deceleration of tree growth, $f(g)$, can be approximated by $\frac{1}{\sqrt{g}}$.

| Parameter name | Estimate | Std. Error | p-value |
| --- | --- | --- | --- |
| $q$ | 0.026 | 4.1284e-04 | <2.2e-16 |
| $\lambda$ | 0.498 | 1.3016e-02 | <2.2e-16 |
| $N_{max}$ | 52.4 | 6.3063e-01 | <2.2e-16 |
| Akaike IC | 52541 | | |
| McFadden's Adj. Pseudo $R^2$ | 86% | | |

**Table 1 | MLE estimations.** The model parameters' estimated values and goodness-of-fit indicators.

**Discussion**

Our findings provide insight into the factors that underlie the emergence and sustainability of online communities. We find that the relationship between activity levels and size in these communities exhibits a three-regime pattern that repeats across platforms and time. Further, we observe a sharp transition between two of the regimes and evidence of the existence of a critical community size. Below that critical size, member activity is largely uncorrelated, and so activity remains low and sporadic. Above that critical size, activity becomes increasingly correlated, and an interactive community emerges. We argue that the regimes' structure and this sharp transition can be explained by a dynamic model of peer-to-peer actions that generates trees of interactions. The model explains the sharp transition as the result of the multiplicative nature of the interactions in which the high-interactivity regime is dominated by the geometric growth of interactions. This geometric growth results from an interplay between community size and the ambient level of responsiveness. In effect, a given level of responsiveness in a group or context defines the characteristic size for the emergence of a sustainable community. A limitation of our findings is that we only observe correlations and are not able, in this non-experimental context, to demonstrate that size actually causes the transition between regimes. Having said that, we find that the model fits the data well, in spite of its relative simplicity, including the observed regimes and the sharp transition patterns. Further, we present indirect and corroborating evidence for the suggested theory. Future work should investigate more complex forms of the model and the implications of some of our simplifying assumptions. Another limitation is while the patterns do mostly replicate across platforms, there are two cases, out of seven, in which one regime is "missing." As explained above, we speculate that this is the result of platform differences. The answer awaits studies with more platforms or studies that investigate the role of platform design. Furthermore, our paper contributes to the computational social science literature; while sharp phase transitions in social systems were hypothesized, mainly by theory[10], the empirical evidence to support these conceptualizations was, so far, lacking. Here, we present the first direct evidence for sharp transitions of collective social behavior. Finally, it is known that within communities, there exists heterogeneity of contribution[3]. This heterogeneity most likely affects the propensity of a community to thrive or fail. Further research should empirically investigate the sources and outcomes of contribution heterogeneity within online communities.

**Methods**

Online discussion groups are constantly created by members in designated online platforms. Generally, a discussion topic initiates with a single seed message posted by a user on the platform. Other users can post replies to that message or to the following messages such that a tree of posts and replies develops. We collected time-stamped group discussion comments at random from context-free platforms such as Tapuz, Goodreads, hi5.com, boards.ie, YouTube, and focal group websites like the Wikipedia article talk pages or the technologically-oriented Reddit. All selected data are available publicly online. Once collected, the records were processed to locate parent–child (directional) links between pairs of comments. In some platforms (e.g., Tapuz), the structure of a discussion page is such that users can choose to respond to a certain post and create a clear thread where each "child" is directly connected to their "parent" response. In other platforms, the child–parent relationship is approximated by either marking the immediate following message, or better still, a user from a preceding comment (the parent comment) may be referenced using hash symbols, similar to the re-tweet mechanism in Twitter. Some platforms (e.g., MediaWiki) may further convert these name mentions to user–page links. These parent–child links collectively thread into a tree-graph of discussions that has measurable depth (max thread path length), volume (number of comments), breadth (number of leaves), community size (number of unique participants), and activity level (number of comments per time unit). These graphs grow with time, and the snapshots we take of the online platform content, therefore, contain the aggregated discussion tree from its initiation until the time of the snapshot. Each platform design is somewhat different, but in order to rule out sample selection and to create consistent data formats, our general rule was to sample from each platform community in a pre-determined time period or at random. For example, the Tapuz data (www.tapuz.co.il/forums) is a collection of all

"communes" (user-generated discussion forums), either active or frozen between the years 2004 and 2016. Using the "random page" function in MediaWiki, 21,000 pages were sampled. In hi5.com, all the discussion topics from 2009 to 2016 were sampled. Similarly, we downloaded YouTube video page data using a third-party tool[1] to randomly sample 10,000 videos. Table 2 provides a descriptive summary of the collected data.

| Dataset | N groups | N users | N posts | Age [yrs] |
|---------|----------|---------|---------|-----------|
| HI5 | 126,468 | 330,936 | 2,568,352 | 7 |
| TAP | 10,122 | 134,747 | 9,986,206 | 12 |
| RED | 14,869 | 764,562 | 10,000,000 | 8 |
| GOODR | 8,040 | 42,345 | 171,411 | 7 |
| WIKI | 17,969 | 30,506 | 154,469 | 15 |
| BRDS | 624,083 | 57,796 | 1,870,566 | 15 |
| YOUT | 8,446 | 1,104,906 | 1,567,073 | 10 |

**Table 2| Summary statistics of the datasets that were used.** HI5 represents data from hi5.com, TAP comes from www.tapuz.co.il/forums, RED are reddit posts from reddit.com/r/datasets/comments/3mg812, GOODR data are from Goodreads www.goodreads.com, WIKI are Wikipedia talk pages, BRDS are www.boards.ie, and YOUT are user comments on YouTube video pages.

---

[1] www.npmjs.com/package/youtube-random-video and tools.digitalmethods.net/netvizz/youtube/mod_video_info.php

# Supporting Information for: Emergence of online communities: Empirical evidence and theory

**Dover and Kelman**

### Supporting Information (SI)

**More patterns of activity vs. size.** Here we show the per-capita activity vs. size relationship for two additional platforms. Figure S1 shows the median per-capita activity for the HI5 dataset. The expected three regimes clearly exist in the curve. Unlike the TAP or BRDS case, the third regime occurs at relatively low mean activity levels. We suspect that the HI5 platform does not encourage longer, chain-like conversations, as, e.g., the TAP and BRDS platforms do. In other words, while discussion trees undergo the sharp transition from Regime I to II, the probability of response decays steeply with tree depth, probably because of the design and context of usage of the platform. The curve in Figure S2 (RED) also exhibits the expected three regimes, with noticeable discrepancies. The first regime occurs only for single-user communities. In that sense, it is similar to the GOODR set. The interpretation here, as it is for GOODR, is that the prevalent responsiveness is already high for micro communities. Further, it seems that the transition into the third regime occurs across a relatively wide range of community sizes. Although out of scope in this paper, we can speculate that this could be the outcome of heterogeneity across communities.

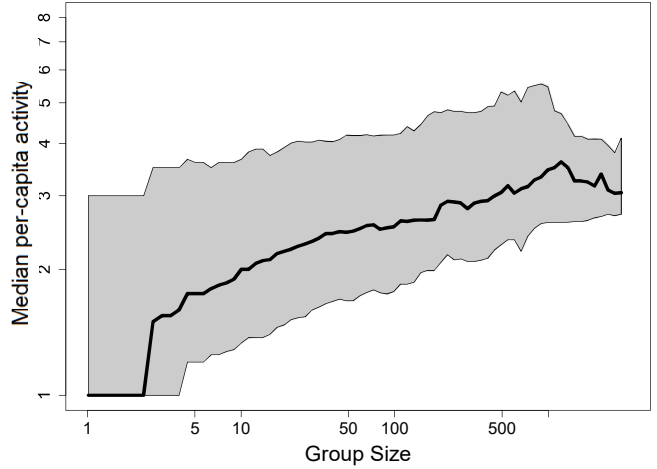

**Fig. S2.** Median group mean activity across a range of community sizes in the RED data. The median activity is the solid thick line, the shaded areas mark the regions between the 25th to the 75th percentiles.

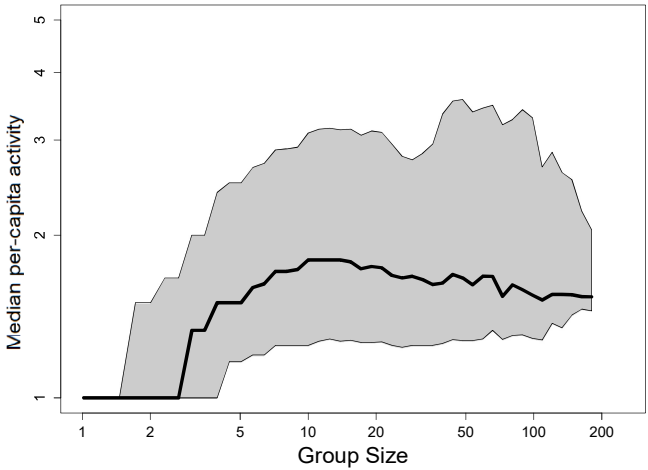

**Fig. S1.** Median group mean activity across a range of community sizes in the HI5 data. The median activity is the solid thick line, the shaded areas mark the regions between the 25th to the 75th percentiles.

**The dependence of discussion tree growth on depth.** We assume, in the model, that the growth of discussion trees, e.g., in equation (4), depends on tree depth. Essentially, this assumption takes into account that as a discussion tree grows longer and deeper, the probability of response to higher-depth messages, decays. One way to model this is by allowing the responsiveness to depend on generation, $g$, i.e., $Q = Q(g)$.

There are several functional forms that can represent this tendency of the response probability to decrease with tree depth that cannot be ruled out, theoretically. With the MLE procedure, we tried several functional forms. Here, we report the two forms that provide the highest level of fit; The exponential decay form: $Q(g) = C \cdot e^{-\lambda \cdot g}$ and the power law decay form: $Q(g) = C \cdot g^{-\lambda}$. The best fit, by far, occurred when we used the power law decay form. Therefore, we use the power law form in the estimation reported in the main part of the paper. Interestingly, the MLE estimation of the power law form, exhibit a value of $\lambda \cong \frac{1}{2}$. In other words, our estimation shows that it seems that the functional form of the rate of deceleration of tree growth, $f(g)$, closely approximates $\frac{1}{\sqrt{g}}$.

**Maximum likelihood estimations and fit.** The purpose of the MLE process is to estimate the parameters of the offspring distribution function $\Phi(\kappa|N, Q(g))$. Notably, the full dynamic picture of discussion tree growth is captured in the offspring distribution function. In our basic scenario, the distribution is homogeneous across tree nodes and solely depends on four parameters: $N$, $N_{\max}$, $q$ and $g$. These are the size of the community, the maximal size of community interaction, the constant community responsiveness and the depth of the node in a discussion tree, respectively. In our case, $N$ is known per community, per node and $g$ is known, per node. The observed random variable, i.e., the number of replies per node, $\kappa$, is also observed. Under the assumptions of the model, the offspring distribution is taken to be binomial. Therefore, the likelihood function can be written as:

$$\mathcal{L}(q, N_{\max}, \lambda) = \prod_{i,j,k} B\big(\mathcal{N}(N_i, N_{\max}), q \cdot f(\lambda, g_{i,j,k})\big) \quad [1]$$

The indices $i, j, k$ are for community, tree and node, respectively. The sample size of the binomial process is $\mathcal{N}(N_i, N_{\max}) = \min\{N_i, N_{\max}\}$, i.e., either community $i$'s size, $N_i$ if $N < N_{max}$ or $N_{max}$ otherwise. The community responsiveness decays with tree depth $g$. The functional form of this decay is expressed by $f(g)$ as explained in the previous section. Here, we write the decay function as $f(\lambda, g_{i,j,k})$ to denote that the depth varies across trees and nodes. Maximum likelihood is obtained by the bounded Broyden–Fletcher–Goldfarb–Shanno (BFGS) approach. The solution is global as we use random initial conditions across a wide range of parameters and runs. The results of the estimation are given in Table 1 in the main paper.

To achieve the fit shown in Figure 6, we used the following process. Using the MLE parameters, we simulated trees as a function of community size. For each community, we simulated 1000 trees and for each community size we created a 1000 simulated communities. The chosen numbers of trees and community sizes did not have a qualitative effect on the results. Using the MLE and model-based simulated communities, we calculated the median per-capita activity as a function of size. The result is shown in Figure 6.